\documentstyle[epsfig]{aipproc}

\begin{document}

\vspace*{-2cm}
\centerline{\small \rm \hfill hep-ph/9711397}
\centerline{\small \rm \hfill ANL-HEP-CP-97-96}
\centerline{\small \rm \hfill November 1997}

\title{Photon Scattering in Muon Collisions\footnote{Talk
given at the Workshop on Physics at the First Muon Collider, Fermilab,
Nov.\ 1997.}}

\author{Michael Klasen\footnote{E-mail: {\tt klasen@hep.anl.gov}.}}
\address{Argonne National Laboratory\thanks{Work supported by the U.S.\ 
         Department of Energy, Division of High Energy Physics, Contracts
         W-31-109-ENG-38 and DEFG05-86-ER-40272.}\\
         High Energy Physics Division\\
         Argonne, Illinois 60439}

\maketitle

\begin{abstract}
We estimate the benefit of muon colliders for photon physics. We
calculate the rate at which photons are emitted from muon beams in different
production mechanisms. Bremsstrahlung is reduced, beamstrahlung disappears,
and laser backscattering suffers from a bad conversion of the incoming to the
outgoing photon beam in addition to requiring very short wavelengths. As a
consequence, the cross sections for jet photoproduction in $\mu p$ and
$\mu^+\mu^-$ collisions are reduced by factors of 2.2 and 5 compared to $ep$ and
$e^+e^-$  machines. However, the cross sections remain sizable and
measurable giving access to the photon and proton parton densities down to $x$
values of $10^{-3}$ to $10^{-4}$.
\end{abstract}

\section*{Introduction}
Muon colliders offer an interesting alternative and complement to future $e^+e^-$
linear colliders. If the considerable design difficulties, in particular in the
multistage cooling, the neutrino radiation problems arising from decaying muons,
and the management of detector backgrounds, can be solved, they can be precisely
tuned to a Higgs or $t\overline{t}$ factory, be ramped in several stages to much
higher energies up to 3 or 4 TeV, and, due to a smaller total size, eventually be
built at lower cost. In addition, an existing high energy proton beam might be
collided with one of the muon beams to give a high energy lepton-proton collider.
The physics interest focuses on measurements of the Higgs and weak gauge boson
properties, top quark physics, supersymmetry with and without R-parity violation,
and other phenomena beyond the Standard Model \cite{Gun97}.

Little attention has been paid in this context to photon initial states and QCD
measurements. This is surprising given the many studies at present and future
$e^+e^-$ colliders, the success of HERA in determining the proton structure and hard
QCD processes, and the significance of QCD background for many of the new physics
processes under consideration. At LEP, ALEPH, DELPHI, and OPAL have studied the
photon structure function $F_2^{\gamma} (x,Q^2)$ in electron-photon scattering
\cite{LEP}, and L3 and OPAL have obtained first measurements of the energy dependence
of the total $\gamma\gamma$ cross section for hadron production \cite{Sol97}.
Whereas LEP1 was dominated by $e^+e^-$ annihilation at the $Z$ pole, $\gamma\gamma$
scattering is important at LEP2 and will be dominant at a future linear $e^+e^-$
collider. This is due to the fact that the annihilation cross section drops
like $\sigma_{l^+l^-}\propto\frac{1}{s}$ or at best like $\frac{\log s}{s}$
($l = e, \mu$), whereas the photon-photon cross section rises like
$\sigma_{\gamma\gamma}\propto \log^3 s$ or even $\propto s$ taking into account the
hadronic structure of the photon. This leads us naturally to the presumption that
photoproduction might not be as negligible in muon collisions as is widely believed.

Literature on muon colliders in general and on photon radiation by muons in
particular is very restricted if not absent. However, two detailed studies on
two-photon physics at $e^+e^-$ linear colliders exist \cite{Dre93,Che94}, where
the authors discuss photon emission and distribution functions, soft and hard
two-photon reactions, and total cross sections. This study follows the lines given
in these references, looks for changes from $e^+e^-$ to $\mu^+\mu^-$, extends them
to $\mu p$ collisions and compares these to $ep$ collisions.
In the following section, we estimate the production rate of photons from muon beams
in three different mechanisms. The results are used in the third section to calculate
the cross section for dijet production as a function of the transverse energy
of the jets and the center-of-mass energy of the muon pair. Finally, we
determine the ranges in $x$ in which the photon and proton structure functions
could be measured at a $\mu^+\mu^-$ and $\mu p$ collider.

\section*{(Quasi-)real Photon Emission}
\begin{figure}[h]
 \begin{center}
  {\unitlength1cm
  \begin{picture}(12,8)
   \epsfig{file=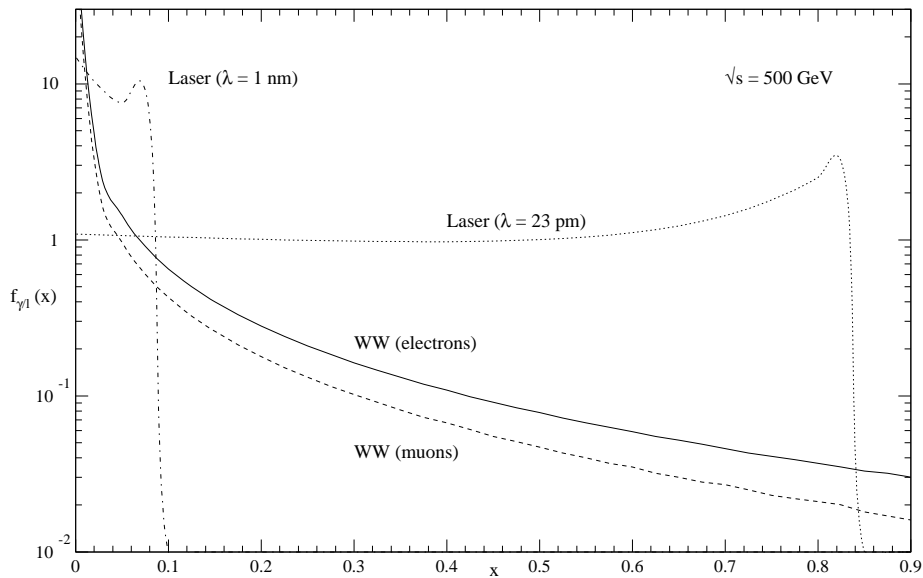,bbllx=520pt,bblly=95pt,bburx=105pt,bbury=710pt,%
           height=12cm,angle=270,clip=}
  \end{picture}}
 \end{center}
 \vspace{10pt}
 \caption{\label{fig1}{\it (Quasi-)real photon emission at a $\sqrt{s} = 500$ GeV
    $\mu^+\mu^-$ collider. We show the Weizs\"acker-Williams (WW) spectra
    for electrons and muons and the spectra for laser backscattering off muons for
    incident photons with wavelengths of 23 pm and 1 nm.}}
\end{figure}

Three mechanisms can contribute to the emission of photons from leptons:
bremsstrahlung, beamstrahlung, and laser backscattering. To begin with,
bremsstrahlung is an approximation of the complete two-photon process
$l^+l^-\rightarrow l^+l^-X$, where $l$ can be an electron or a muon, and $X$ is a
hadronic final state in our case. The photons are radiated from the
leptons in the scattering process, and their spectrum can be expressed through the
usual Weizs\"acker-Williams or Equivalent Photon Approximation \cite{Fri93}
\begin{equation}
 f_{\gamma/l}^{\rm brems} (x) = \frac{\alpha}{2\pi} \left[ \frac{1+(1-x)^2}{x}
 \log\frac{Q_{\max}^2}{Q_{\min}^2}+2m^2_lx\left(\frac{1}{Q_{\max}^2}-\frac{1}
 {Q_{\min}^2}\right)\right], \label{eq1}
\end{equation}
where $x = \frac{E_{\gamma}}{E}$ is the fraction of the photon energy $E_{\gamma}$
of the lepton beam energy $E$, and where the virtuality of the photon
\begin{equation}
 Q^2 = \frac{m^2_lx^2}{1-x}+E^2(1-x)\theta^2 + {\cal O} \left(E^2\theta^2,
 m^2_l\theta^2, \frac{m^2_l}{E^2}\right) \label{eq2}
\end{equation}
is always smaller than
\begin{equation}
 Q^2 = - (p_l-p_{l'})^2 = 2 EE'(1-\cos\theta) < 4EE' = 4E^2(1-x).
\end{equation}
This upper limit on $Q^2$ is used if no information on the scattered lepton is
available. For a safe factorization of the lepton tensor and phase space, it is
preferred to anti-tag the outgoing lepton and use the maximum scattering
angle $\theta$ in Eq.\ (\ref{eq2}). Consequently, the photon density will depend on
this maximum scattering angle. This is shown in Table \ref{tab1} for a
$\sqrt{s} = 500$ GeV $e^+e^-$ collider.
\begin{table}[h!]
\caption{Dependence of the photon density on the maximum scattering angle.}
\begin{tabular}{|c|c|}
\hline
$\theta_{\max}$ & $f_{\gamma/e}^{\rm brems} (x = 0.5)$ \\
\hline
\hline
 ---    & 0.078 \\
10 mrad & 0.047 \\
20 mrad & 0.051 \\
30 mrad & 0.053 \\
\hline
\end{tabular}
\label{tab1}
\end{table}
The variation in this case can reach up to 40\%. Which changes occur in the
Weizs\"acker-Williams spectrum for muon beams? Eq.\ (\ref{eq1}) depends explicitly
on the lepton mass $m_l$ in its non-logarithmic contribution as well as implicitly
through the minimal photon virtuality $Q^2_{\min} = \frac{m^2_lx^2}{1-x}$. Taking
into account only the leading logarithmic contribution, we expect a reduction
by a factor of
\begin{equation}
 \frac{\log\frac{Q_{\max}^2(1-x)}{m_{\mu}^2x^2}}
      {\log\frac{Q_{\max}^2(1-x)}{m_{e}^2x^2}} =
 \frac{-\log m_{\mu}^2}{-\log m_e^2} \simeq
 \frac{1}{3.38},
\end{equation}
where we have taken $Q_{\max}^2 = 1$ GeV$^2$ and $1-x = x^2$ or $x= 0.618$ for
simplicity. If one considers a muon collider with a photon spectrum on either
side, one expects the cross section to be reduced by a factor of
$\left(\frac{1}{3.38}\right)^2 \simeq \frac{1}{11.4}$.
In Figure \ref{fig1}, we compare the Weizs\"acker-Williams (WW) spectra
of electrons and muons as a function of $x$. We have not assumed any anti-tagging
conditions on the scattered lepton. The shape of the muon spectrum is
basically unchanged with respect to the electron spectrum, and the normalization
drops by a factor of two. This is in qualitative agreement with our na\"{\i}ve
estimate above.

A second contribution to photon emission comes from beamstrahlung. The particles
in one bunch experience rapid acceleration when they enter the electromagnetic
field of the opposite bunch. Then the scattering amplitudes between particles
within the characteristic length add coherently. This can involve up to $10^6$
particles. The intensity and spectrum of the beamstrahlung depend sensitively on
the size and shape of the bunches and thus on the machine parameters. It is known
that for particular $e^+e^-$ linear collider designs beamstrahlung can be more
important than bremsstrahlung over a wide range in $x$ \cite{Dre93}. Within a
semiclassical calculation and for a Gaussian longitudinal bunch profile, Chen
et al. find
\begin{equation}
 f_{\gamma/l}^{\rm beam} (x) = \frac{1}{\Gamma\left(\frac{1}{3}\right)}
 \left[ \frac{2}{3{\cal Y}} \right]^{1/3} x^{-2/3}(1-x)^{-1/3}
 \exp\left[-\frac{2x}{3{\cal Y} (1-x)} \right] G(x),
 \label{eq3}
\end{equation}
where the function $G(x)$ is of order 1 and can found in \cite{Che94}.
The beamstrahlung intensity is controlled by the effective beamstrahlung parameter
\begin{equation}
 {\cal Y} = \frac{5 r_l^2 E N}{6\alpha \sigma_z(\sigma_x+\sigma_y)m_l}
\end{equation}
mostly through the exponential function.
\begin{equation}
 r_l = \frac{e^2}{4\pi\epsilon_0m_lc^2}
\end{equation}
is the ``classical radius'' of each of the $N$ leptons with mass $m_l$ in
a bunch of length $\sigma_z$ and transverse dimensions $\sigma_x$ and $\sigma_y$.
$E$ is the beam energy, and $\alpha=\frac{e^2}{4\pi}$ is the fine structure
constant. The beamstrahlung parameter ${\cal Y}$ depends explicitly and
implicitly through the classical lepton radius on the lepton mass $m_l$.
We therefore expect the beamstrahlung parameter for muon beams to be reduced by
\begin{equation}
 \frac{{\cal Y}_{\mu}}{{\cal Y}_e} = \frac{m_e^3}{m_{\mu}^3} \simeq 8*10^{-6},
\end{equation}
i.e. by three powers of the mass ratio of electrons to muons. Since the
beamstrahlung intensity will be suppressed by this factor exponentially, we
do not expect any beamstrahlung for muon beams and in fact do not see any
in a numerical simulation using the formula of Eq.\ (\ref{eq3}).

$e^+e^-$ colliders can also be run in a mode dedicated to $\gamma\gamma$
collisions using backscattering of laser light off the incident electron
and positron beams. One considers even dedicated interaction regions.
For unpolarized lepton beams, the spectrum of the backscattered photons
depends only on the beam energy, on the laser wavelength, and --- this is
crucial for muon beams --- on the beam particle mass.  Ginzburg et al.
determined this spectrum to be \cite{Gin83}
\begin{equation}
 f_{\gamma/l}^{\rm laser} (x) = \frac{1}{N}\left[ 1-x+\frac{1}{1-x}
 -\frac{4x}{X(1-x)}+\frac{4x^2}{X^2(1-x)^2}\right],
\end{equation}
where the normalization is
\begin{equation}
 N = \left[ 1-\frac{4}{X}-\frac{8}{X^2}\right]
 \log (1+X) +\frac{1}{2}+\frac{8}{X}-\frac{1}{2(1+X)^2}.
\end{equation}
As already stated above, the crucial parameter
\begin{equation}
 X = \frac{4EE_{\gamma}^{\rm in}}{m_l^2}
\end{equation}
in this case depends on the beam energy $E$, on the incident photon energy
$E_{\gamma}^{\rm in}= \frac{hc}{\lambda}$, and on the lepton mass $m_l$.
Telnov estimates the optimal value of $X$ from the threshold of the $e^+e^-$
pair creation in photon collisions being
\begin{equation}
 E_{\gamma}^{\rm out} E_{\gamma}^{\rm in} = \frac{X^2m_l^2}{4(X+1)} > m_e^2.
\end{equation}
For electron beams, $m_l = m_e$ cancels in the equation above leading to an
optimal value of $X^{\rm opt}_e=2(1+\sqrt{2})$. For muon beams, we find
\begin{equation}
 X^{\rm opt}_{\mu} = 2\frac{m_e^2}{m_{\mu}^2}
 \left(1+\sqrt{1+\frac{m_{\mu}^2}{m_e^2}}\right) \simeq 9.72*10^{-3}.
\end{equation}
In Table \ref{tab2} we calculate the optimal laser wavelength as a function 
\begin{table}[h!]
\caption{Dependence of the laser wavelength on the center-of-mass energy.}
\begin{tabular}{|c|c|c|c|c|}
\hline
$\sqrt{s}$/GeV & $E_{e,\mu}$/GeV & $\lambda_{\gamma}^e$/nm & $\lambda_{\gamma}^
{\mu,{\rm equiv}}$/pm & $\lambda_{\gamma}^{\mu,{\rm opt}}$/nm \\
\hline
\hline
100 &  50 & 197 &  4.6 & 2.29 \\
200 & 100 & 393 &  9.2 & 4.57 \\
350 & 175 & 688 & 16.1 & 8.00 \\
500 & 250 & 983 & 23.0 & 11.4 \\
\hline
\end{tabular}
\label{tab2}
\end{table}
of the center-of-mass energy $\sqrt{s}$ of the lepton collider or equivalently
the lepton beam energy $E_{e,\mu}$. For electron beams, the optimal wavelengths
lie in the region of visible light (197 - 983 nm) and can be provided by current
laser technology. To obtain the same backscattered photon spectrum as with
electron beams, much shorter wavelengths are needed at a muon collider (4.6 -
23.0 pm). The spectrum for $\lambda = 23$ pm is shown in Figure \ref{fig1}.
This spectrum does, however, not take into account the bad conversion of the
incoming to the outgoing photon beam due to enhanced $e^+e^-$ pair creation.
If one chooses the optimal value of $X_{\mu}^{\rm opt} = 9.72*10^{-3}$ for
100 \% conversion and no $e^+e^-$ pair production, one finds incident laser
wavelengths of 2.29 - 11.4 nm. However, a laser beam of 11.4 nm produces
backscattered photons at much too low energies which do not appear on
the spectrum of Figure \ref{fig1} at all. For illustration, we show the spectrum
for $\lambda = 1$ nm, where the spectrum is still concentrated at low energies,
but at least visible. However, the conversion will be bad, and these short
wavelengths can only be obtained with new laser technology like free electron
lasers.

\section*{QCD and $\gamma\gamma$ Scattering}

\begin{figure}[p]
 \begin{center}
  {\unitlength1cm
  \begin{picture}(12,8)
   \epsfig{file=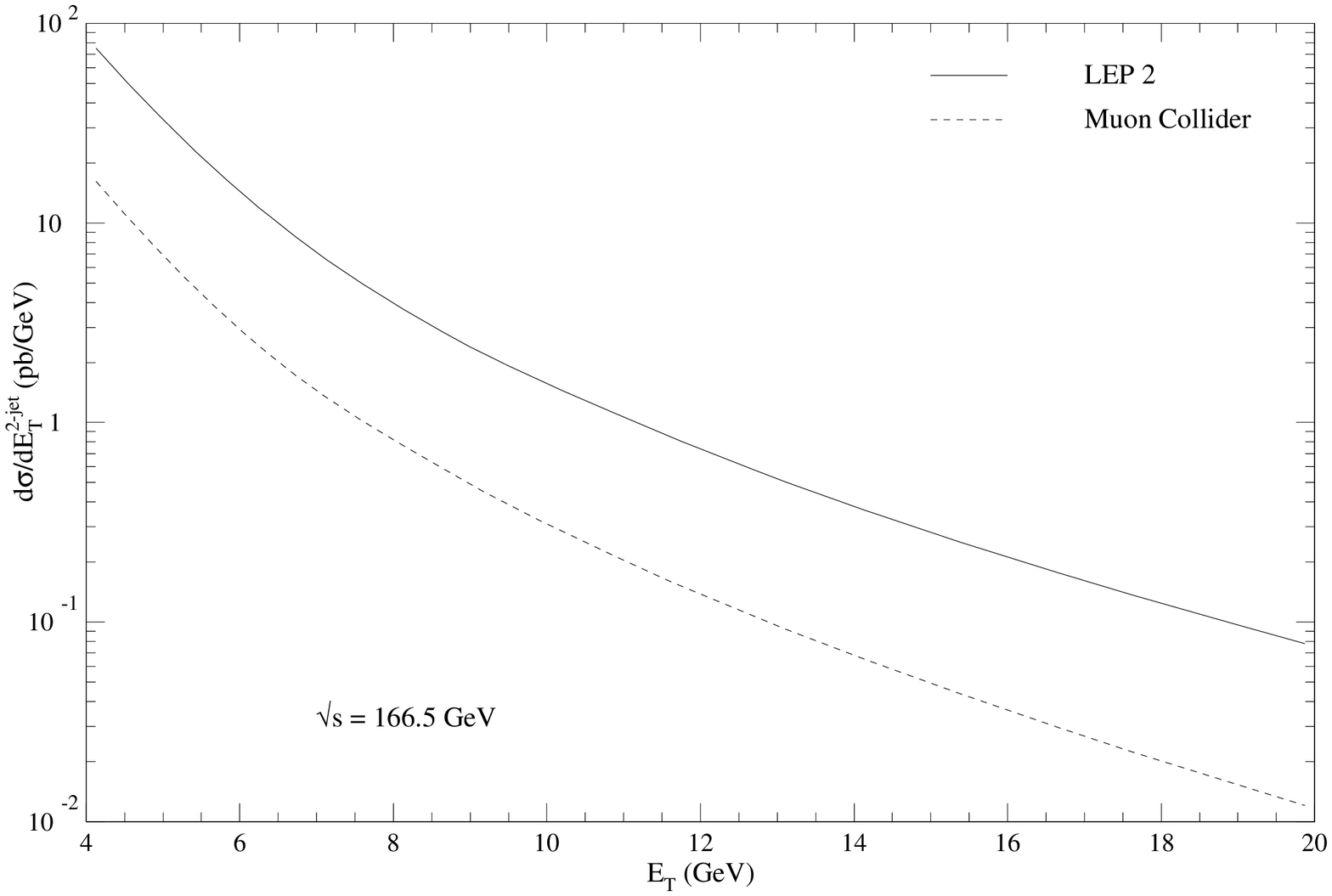,bbllx=520pt,bblly=95pt,bburx=105pt,bbury=710pt,%
           height=12cm,angle=270,clip=}
  \end{picture}}
 \end{center}
 \vspace{10pt}
 \caption{\label{fig2}{\it Comparison of the differential dijet cross section
 at LEP2 and a muon collider of the same center-of-mass energy as a function
 of the transverse energy $E_T$.}}
 \begin{center}
  {\unitlength1cm
  \begin{picture}(12,8)
   \epsfig{file=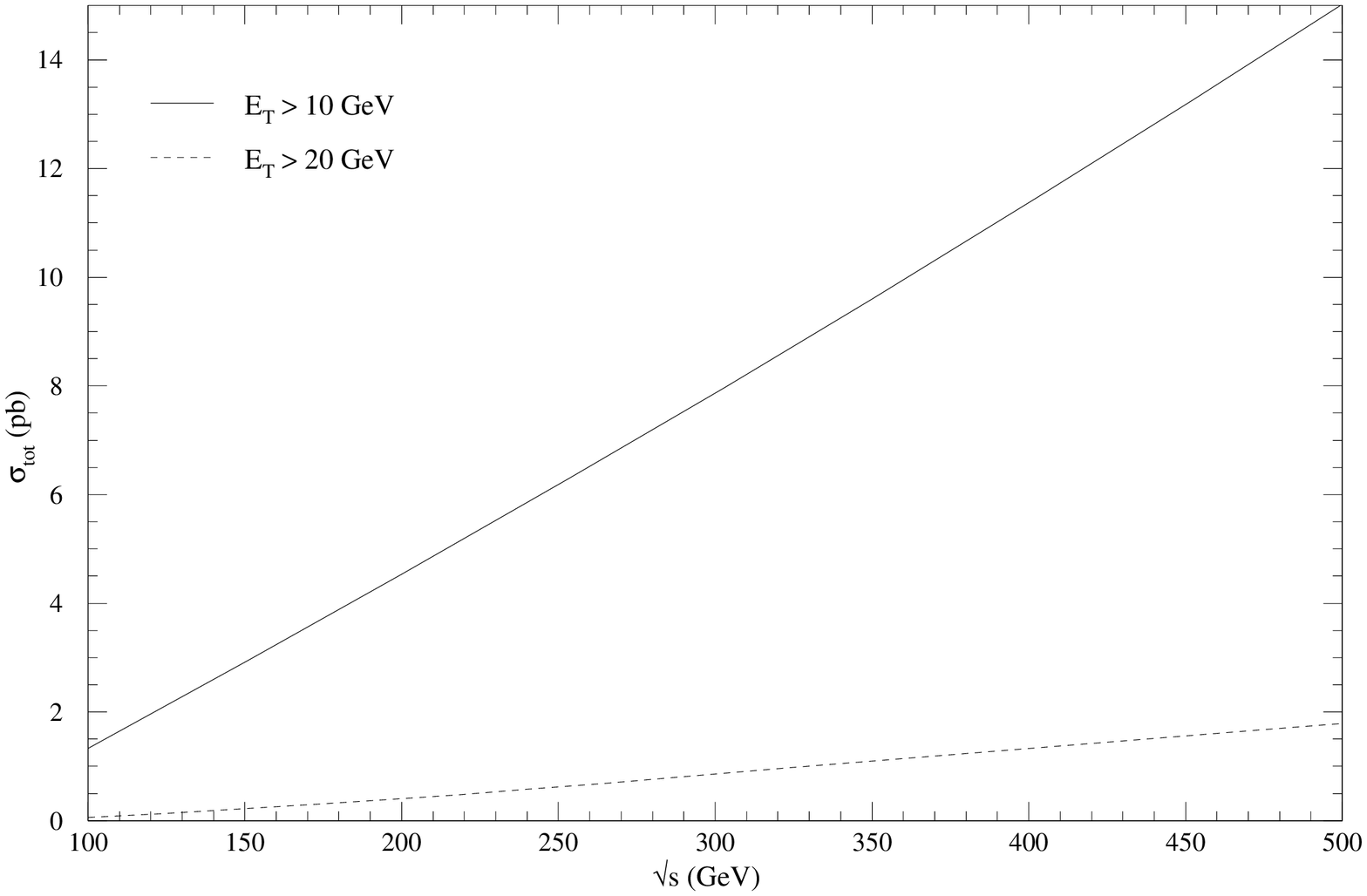,bbllx=520pt,bblly=95pt,bburx=105pt,bbury=710pt,%
           height=12cm,angle=270,clip=}
  \end{picture}}
 \end{center}
 \vspace{10pt}
 \caption{\label{fig3}{\it Center-of-mass energy dependence of the total dijet cross
 section for two thresholds of the transverse energy $E_T$.}}
\end{figure}

In this section we will restrict ourselves to the reduced bremsstrahlung spectrum
of muon beams and apply it to QCD and $\gamma\gamma$ scattering. As an example,
we compare the differential dijet cross section d$\sigma$/d$E_T^{\rm 2-jet}$ at LEP2
and a muon collider of the same center-of-mass energy $\sqrt{s} = 166.5$ GeV as a
function of the transverse energy $E_T$ in Figure \ref{fig2}.
This cross section has recently been measured by OPAL for 3 GeV $< E_T < 20$ GeV,
$-2 < \eta_{1,2} < 2$, and a maximum electron scattering angle of $\theta = 33$
mrad using an integrated luminosity of about 20 pb$^{-1}$ \cite{Sol97}. It drops
from 133 to 0.16 pb/GeV from the first to the last $E_T$-bin. The prediction for
a muon collider is reduced by factors of 4.64 to 6.47 in this $E_T$-range. The
cross section is still large enough to be measured with good precision.

It is also interesting to look at the dependence of the total dijet cross section
on the center-of-mass energy. The result is shown in Figure \ref{fig3} for two
cuts on the transverse energy at 10 and 20 GeV. We observe a
linear rise of the cross section as we expect when taking into account the hadronic
structure of the photon. Photoproduction of jets will therefore increase
linearly in importance with the energy at which a muon collider is operated.

\section*{Parton Density Measurements}

\begin{figure}[p]
 \begin{center}
  {\unitlength1cm
  \begin{picture}(12,8)
   \epsfig{file=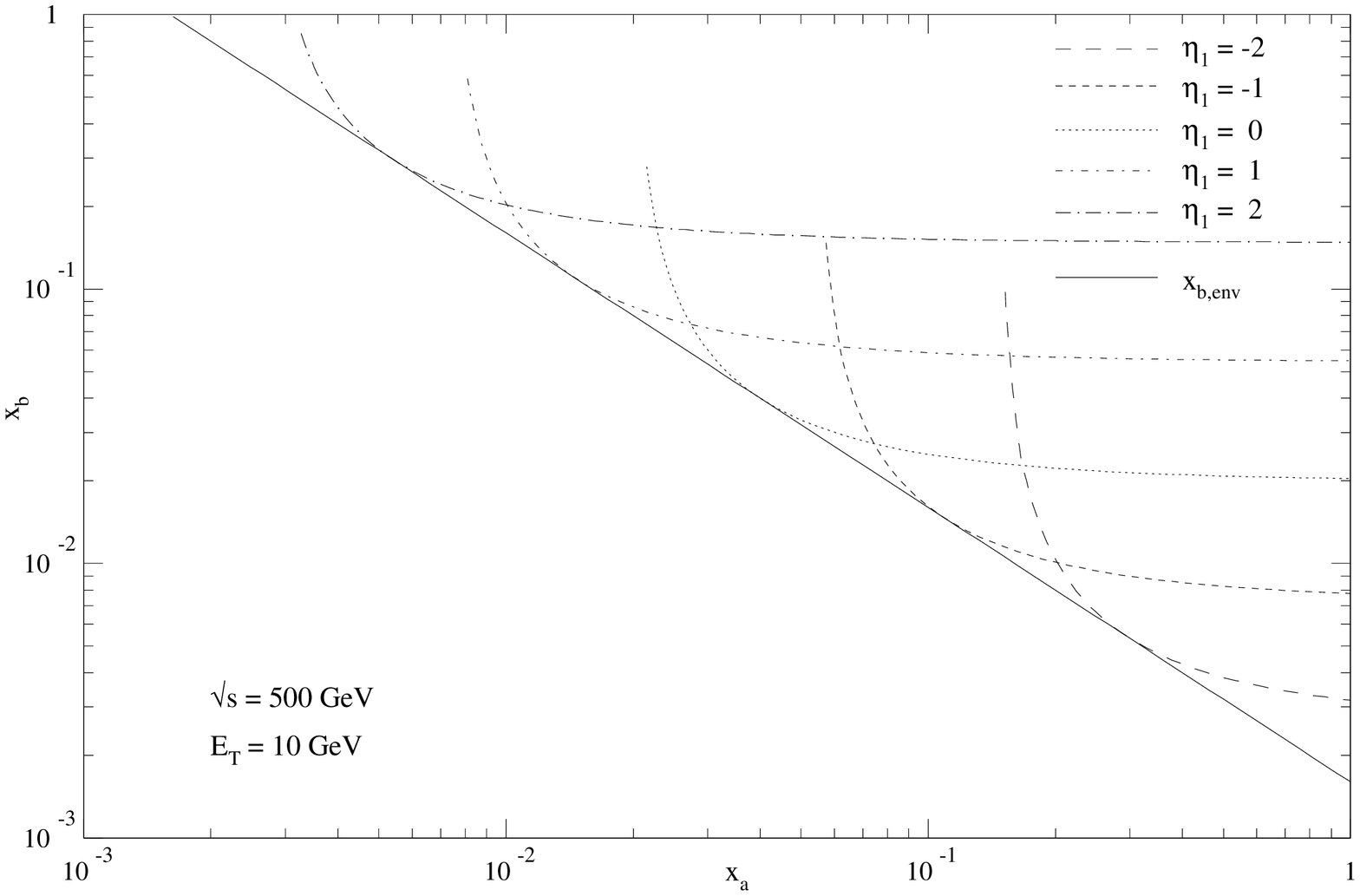,bbllx=520pt,bblly=95pt,bburx=105pt,bbury=710pt,%
           height=12cm,angle=270,clip=}
  \end{picture}}
 \end{center}
 \vspace{10pt}
 \caption{\label{fig4}{\it Parton density regions in the $x_a - x_b$ plane that can 
 be accessed at a $\sqrt{s} = 500$ GeV muon collider for different rapidities $\eta_1$
 and $E_T = 10$ GeV.}}
 \begin{center}
  {\unitlength1cm
  \begin{picture}(12,8)
   \epsfig{file=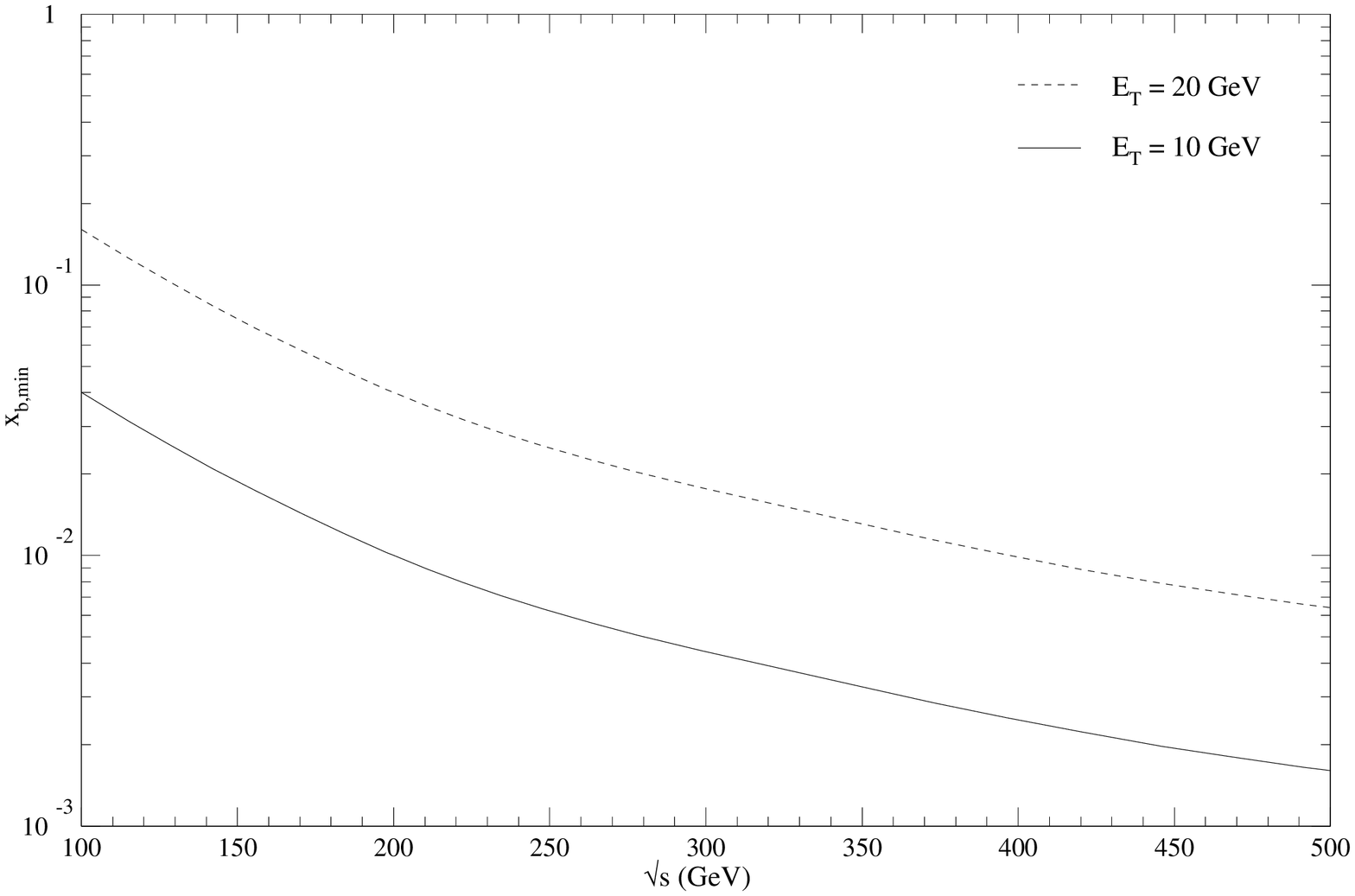,bbllx=520pt,bblly=95pt,bburx=105pt,bbury=710pt,%
           height=12cm,angle=270,clip=}
  \end{picture}}
 \end{center}
 \vspace{10pt}
 \caption{\label{fig5}{\it Center-of-mass energy dependence of the minimal $x_b$ value
 for two different transverse energies $E_T = 10$ and $20$ GeV.}}
\end{figure}

We will now investigate the prospects for measuring the photon and proton
parton densities at $\mu^+\mu^-$ and $\mu p$ colliders in photoproduction
processes. The $x$-ranges that are accessible can be calculated from four-momentum
conservation and therefore depend completely on the kinematics of the initial and
final states. The momentum fraction $x_b$ of a parton $b$ in the correspondent
parent particle (muon or proton) is given by
\begin{equation}
 x_b = \frac{x_aE_aE_Te^{\eta_1}}{2x_aE_aE_b-E_bE_Te^{-\eta_1}}.
\end{equation}
It depends on the momentum fraction $x_a$ of parton $a$ in its parent particle, on
the beam energies $E_{a,b}$, on the transverse momentum of the outgoing particles $E_T$,
and on the rapidity $\eta_1$ of one of these. The second rapidity $\eta_2$ is linearly
dependent on these variables. We have plotted this function in Figure \ref{fig4}
for a muon collider with $\sqrt{s} = 500$ GeV and different rapidities $\eta_1 \in
[-2;2]$. This range corresponds to the fairly limited rapidity coverage of a detector
at muon colliders due to massive shielding of decay electrons from the muon beams.
The transverse energy $E_T = 10$ GeV has been chosen in such a way that one may hope
for a sufficient suppression of non-perturbative effects and a good determination of
the energy scale which are indispensable for a safe extraction of the parton densities.
Figure \ref{fig4} also shows the envelope
\begin{equation}
 x_{b,{\rm env}} = \frac{E_T^2}{x_aE_aE_b},
\end{equation}
which gives the absolute limits on $x_{a,b}$ independent of the rapidity
$\eta_1$. We find that for the kinematical conditions described above, rather low
values of $x_{a,b} > 1.6*10^{-3}$ can be reached where little is known about the
parton densities (in particular of the gluon) in the photon.

Since it is not clear at which energy a muon collider will eventually operate, we
also consider the center-of-mass energy dependence of the minimal $x$ value that can
be reached in Figure \ref{fig5}. For $E_T = 10$ (20) GeV, one can only go down to 0.04
(0.16) for a 100 GeV collider. Higher energies are therefore a clear advantage for
photoproduction both for higher cross sections and for larger $x$-ranges.

\begin{figure}[h]
 \begin{center}
  {\unitlength1cm
  \begin{picture}(12,8)
   \epsfig{file=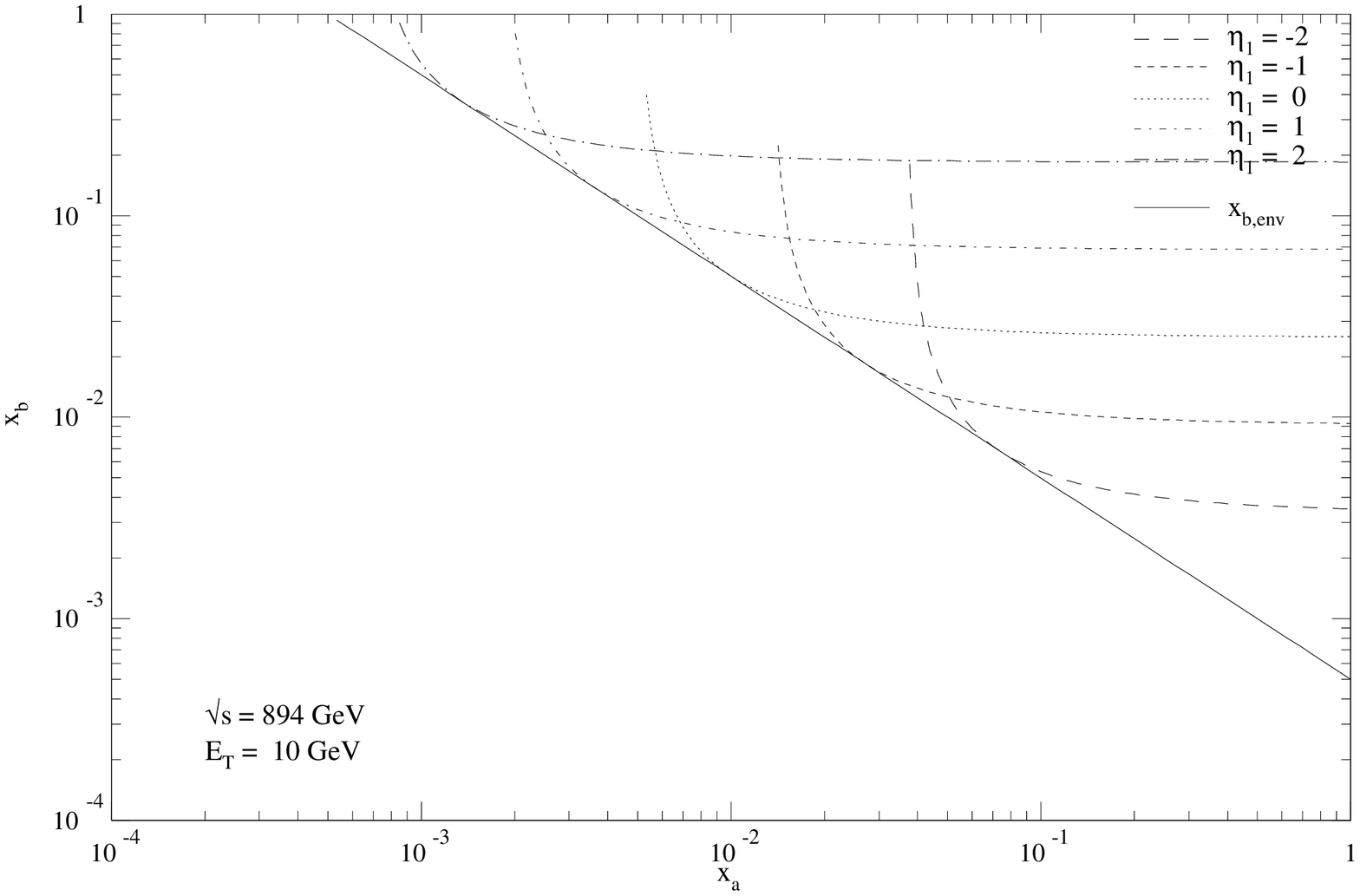,bbllx=520pt,bblly=95pt,bburx=105pt,bbury=710pt,%
           height=12cm,angle=270,clip=}
  \end{picture}}
 \end{center}
 \vspace{10pt}
 \caption{\label{fig6}{\it Parton density regions in the $x_a - x_b$ plane that can 
 be accessed at a $\sqrt{s} = 894$ GeV muon-proton collider for different rapidities
 $\eta_1$ and $E_T = 10$ GeV.}}
\end{figure}

Finally, we repeat the above analysis for a 200 GeV x 1000 GeV muon-proton collider.
The result is shown in Figure \ref{fig6}.
We find that the parton densities in the photon and proton could be measured down to
values of ${\cal O} (5*10^{-4})$.

\section*{Conclusions}
We have studied several aspects of photon physics at muon colliders. Photon
emission due to bremsstrahlung is found to be reduced, there is no beamstrahlung,
and laser backscattering appears to be difficult. However, QCD cross sections
like photoproduction of dijets remain sizable and increase linearly with the
center-of-mass energy. Photon and proton parton densities could be measured
down to $x$ values of $10^{-3}$ to $10^{-4}$.

\end{document}